\documentclass[9pt]{article}
\usepackage{spconf,amsmath,graphicx}
\usepackage[bookmarks,colorlinks,citecolor=green]{hyperref}
\usepackage{acronym}
\usepackage{amssymb}
\usepackage{cite}
\usepackage{color}
\usepackage{xurl}
\usepackage[ruled,linesnumbered]{algorithm2e}  
\SetKwInput{KwData}{\textbf{Init}} 
\let\oldnl\nl
\newcommand{\nonl}{\renewcommand{\nl}{\let\nl\oldnl}}
\usepackage{algpseudocode} 
\usepackage[bottom]{footmisc}

\acrodef{eeg}[EEG]{electroencephalogram}
\acrodef{hmm}[HMM]{hidden Markov model}
\acrodef{map}[MAP]{maximum a-posteriori probability}
\acrodef{dl}[DL]{deep learning}
\acrodef{dnn}[DNN]{deep neural network}
\acrodef{cnn}[CNN]{convolutional neural network}
\acrodef{dwt}[DWT]{discrete wavelet transform}
\acrodef{rnn}[RNN]{recurrent neural network}
\acrodef{flops}[FLOPs]{floating point operations}
\acrodef{lstm}[LSTM]{Long-Short Term Memory}
\acrodef{mi}[MI]{mutual information}
\acrodef{kl}[KL]{Kullback-Leibler}
\acrodef{mical}[MICAL]{Mutual Information-based CNN-Aided Learned factor graphs}

\title{CNN-Aided Factor Graphs with Estimated\\ Mutual Information Features for Seizure Detection}
\name{Bahareh Salafian\textsuperscript{1},Eyal Fishel Ben-Knaan\textsuperscript{2},Nir Shlezinger\textsuperscript{2}, Sandrine de Ribaupierre\textsuperscript{3}, Nariman Farsad\textsuperscript{1}}

\address{\textsuperscript{1}Department of Computer Science, Ryerson University,\\
\textit {\{bsalafian,nfarsad\}@ryerson.ca}\\
\textsuperscript{2}School of Electrical and Computer Engineering, Ben-Gurion University of the Negev,\\
\textit{eyalfish@post.bgu.ac.il}, \textit{nirshl@bgu.ac.il}
\\
\textsuperscript{3}School of Biomedical Engineering, University of Western Ontario,\\
\textit{sderibau@uwo.ca}}

\begin{document}
\ninept
\maketitle
\begin{abstract}
We propose a convolutional neural network (CNN) aided factor graphs assisted by mutual information features estimated by a neural network for seizure detection. Specifically, we use neural mutual information estimation to evaluate the correlation between different electroencephalogram (EEG) channels as features. We then use a 1D-CNN to extract extra features from the EEG signals and use both features to estimate the probability of a seizure event.~Finally, learned factor graphs are employed to capture the temporal correlation in the signal. Both sets of features from the neural mutual estimation and the 1D-CNN are used to learn the factor nodes. We show that the proposed method achieves state-of-the-art performance using 6-fold leave-four-patients-out cross-validation.   
\end{abstract}
\vspace{-0.2cm}
\section{Introduction}
\label{sec:intro}
Epilepsy is a highly common neurological disorder, causing recurrent episodes of the involuntary movement known as epileptic seizures~\cite{litt_prediction_2002}. Based on the place in the brain where seizure starts and the intensity of the abnormal signals, patients with epilepsy may suffer from different symptoms such as auras, repetitive muscle contraction, and loss of consciousness.~\cite{park_epileptic_2018}. Epileptic seizure severely affects the patient’s quality of life and can have other social and economic impacts; for instance, some activities, including swimming, bathing, and climbing a ladder, become dangerous as a seizure during that activity might result in unpredictable injuries and even death. 
Therefore, early detection of epilepsy can notably improve the patient's quality of life. A leading tool to diagnose seizure is based on~\ac{eeg} monitoring, being economical, portable, and non-invasive~\cite{subasi_epileptic_2019}. However, the review of \ac{eeg} recordings is a time-consuming expert-dependent process due to contamination by physiological and non-physiological resources~\cite{golmohammadi_deep_2017}, and similarity of epileptic spikes to normal \ac{eeg} waveforms. 

The challenges associated with \ac{eeg} monitoring gave rise to a growing interest in machine learning aided automatic seizure detection. 
A common approach is to train a model, typically a \ac{cnn}, applied to features extracted from the Wavelet or Fourier transform of the signal~\cite{slimen_eeg_2020,ahmad_prediction_2020,raghu_automated_2020,li_seizure_2021,sharan_epileptic_2020,jana_1d-cnn-spectrogram_2020}, typically involving careful feature engineering. Other seizure detection methods process the raw \ac{eeg} signals directly. These include the application of \ac{cnn}s~\cite{khalilpour_application_2020, boonyakitanont_comparison_2019} to the segmented \ac{eeg}s (e.g., 4-second blocks), providing instantaneous prediction without exploiting temporal correlation between blocks. Prior works have also considered \ac{cnn}-\ac{rnn} architectures to capture temporal correlation~\cite{roy18GRU, liang_scalp_2020} that lead to high computational complexity during training.
The challenges associated with previous works motivate the formulation of a reliable automatic seizure detection algorithm which generalizes to different patients, benefits from both temporal and inter-channel correlation, and is computationally efficient facilitating its application in real-time. 

In this work, we propose a data-driven automatic seizure detection system coined \ac{mical}. \ac{mical} combines computationally efficient 1D \acp{cnn} with principled methods for benefiting from temporal and inter-channel correlation. Following \cite{salafian2021efficient}, we exploit the temporal correlation by imposing a Markovian model on the latent seizure activity \cite{lee_classification_2018}, using the \ac{cnn} output not as seizure estimates, but as messages conveyed as a form of learned factor graph inference \cite{shlezinger2020inference,knobelreiter2020belief,shlezinger2020data}. We expand our previous work~\cite{salafian2021efficient}, to exploit the inter-channel correlation during the seizure by estimating the \ac{mi} between each pair of \ac{eeg} channels through a neural \ac{mi} estimator. To the best of our knowledge, this is the first time \ac{mi} has been used as the feature for seizure detection. The \ac{mi} features along with features learned by the 1D-CNN are then used for learning the factor nodes for factor graph inference. Our numerical evaluations, which use the CHB-MIT dataset \cite{goldberger_physiobank_2000}, demonstrate how each of the ingredients combined in \ac{mical} contributes to its reliability, allowing it to achieve improved accuracy and generalization performance compared to previous algorithms. 

The rest of this paper is organized as follows. In Section~\ref{sec:Model} we describe the problem statement and review necessary preliminaries. Then, in Section~\ref{sec:Method} we describe the proposed \ac{mical} algorithm; Section~\ref{sec:Sims} presents a numerical study, while  Section~\ref{sec:Conclusions} provides concluding remarks.
\section{Problem Statement and Preliminaries}
\label{sec:Model}
\vspace{-2mm}
\subsection{Seizure Detection Problem}
\label{subsec:Problem}
In this paper, seizure detection refers to the identification and localization of the ictal (i.e., the seizure) time intervals from EEG recordings of patients with epilepsy~\cite{emmady_eeg_2020}. To formulate this mathematically, let $\boldsymbol{X}=\{\boldsymbol{X}_{1},\boldsymbol{X}_{2},\cdots,\boldsymbol{X}_{N} \}$ be the EEG recordings of a patient, where $N$ represents the number of channels. Each measured channel $\boldsymbol{X}_i$ is comprised of $n$ consecutive blocks, e.g., blocks of 1-second recordings, and we write $\boldsymbol{X}_i=[\boldsymbol{x}^{(i)}_{t_1},\boldsymbol{x}^{(i)}_{t_2},\cdots,\boldsymbol{x}^{(i)}_{t_n}]$, where $\boldsymbol{x}^{(i)}_{t}$ is the signal corresponding to the $i$-th \ac{eeg} channel during the $t$-th block. The seizure state for each block is represented as a binary vector $\boldsymbol{s}=[s_{t_1},\ldots s_{t_n}]$, where $s_t\in\{0,1\}$ models whether or not a seizure occurs in the $t$-th block. Our goal is to design a system which maps the \ac{eeg} recordings $\boldsymbol{X}$ into an estimate of $\boldsymbol{s}$, which is equivalent to finding the time indices where seizure occurs. 

To model the relationship between the \ac{eeg} signals $\boldsymbol{X}$ and the seizure states $\boldsymbol{s}$, it is needed to consider both inter-channel correlation as well as temporal correlation that the recordings exhibit. The former stems from the fact that when the seizure starts, the epileptic activity propagates to other areas in the brain~\cite{quintero2016new} which affects the patterns of other channel recordings~\cite{jemal2021study}. 
This manifests high dependence between different channels, i.e., between $\boldsymbol{x}^{(i)}_{t}$ and $\boldsymbol{x}^{(j)}_{t}$, when $t$ is at the beginning and during ictal phase. Temporal correlation results from the fact that seizures typically span multiple recording blocks, and thus the probability of observing a seizure at time instance $t$ depends on the presence of a seizure in the previous block, such that the entries of $\boldsymbol{s}$ can be approximated by a Markovian structure \cite{lee_classification_2018}. Our proposed solution, detailed in Section~\ref{sec:Method}, exploits this statistical structure using 
factor graphs.

\subsection{Factor Graph Inference}
\label{subsec:Factor}
Factor graphs are a representation of factorizable multi-variable functions, such as probability distributions, as a bipartite graph. These graphical models facilitate inference at reduced complexity via message passing algorithms, such as the sum-product methods \cite{loeliger2004introduction}. Consider an observed sequence $\boldsymbol{Y} = [\boldsymbol{y}_1,\ldots, \boldsymbol{y}_n]$ encapsulating a latent state sequence $\boldsymbol{s}=[s_1,\ldots,s_n]$ whose entries take values in a finite set $\mathcal{S}$, as a form of a \ac{hmm}. In such cases, the joint distribution of $\boldsymbol{y}, \boldsymbol{s}$ obeys 
\begin{equation}
    \label{eqn:factor}
	P(\boldsymbol{s},\boldsymbol{y}) =  \prod_{k = 1}^{n} P(s_{k}|s_{k-1}) P(\boldsymbol{y}_{k} | s_{k}),
\end{equation}
which can be represented as factor graph with variable nodes $\{s_k\}_{k=1}^n$ and function nodes $\{f_k\}_{k=1}^n$, where $f_k(s_k, s_{k-1}):= P(s_{k}|s_{k-1}) P(\boldsymbol{y}_{k} | s_{k})$.

The factor graph representation allows one to compute the \ac{map} decision rule with a complexity that only grows linearly with $n$ as opposed to exponentially with $n$. This is achieved by evaluating the marginal distribution $P({s}_k,\boldsymbol{y})$ for each $k \in \{1,\ldots n\}$ via  message passing over the factor graph. In this case, the forward messages are recursively updated via
\begin{equation}
\label{eqn:ForwardPass}
\mu_{f_{k}\rightarrow s_{k}}(s_{k})=\sum_{\{s_{1},\cdots,s_{i-1} \}}\prod_{j = 1}^{k}f_{j}(s_{j},s_{j-1}),
\end{equation}
and the backward messages via
\begin{equation}
\label{eqn:BackwardPass}
\mu_{f_{k+1}\rightarrow s_{k}}(s_{k})=\sum_{\{s_{k+1},\cdots,s_{n} \}}\prod_{j = k+1}^{n}f_{j}(s_{j},s_{j-1}).
\end{equation}
Then, the desired marginal distribution, which is maximized by the \ac{map} rule, is given by
\begin{equation}
\label{eqn:msgs}
    P(s_{k},\boldsymbol{y}) = \mu_{f_{k}\rightarrow s_{k}}(s_{k})\cdot \mu_{f_{k+1}\rightarrow s_{k}}(s_{k}).
\end{equation}
Intuitively \eqref{eqn:ForwardPass}-\eqref{eqn:BackwardPass} are interpreted as an aggregate of neighboring information. Once all neighbors have communicated (i.e., messages have propagated the entirety of the graph) the product of the forwards and backward messages determines the marginal probability.

\section{MICAL Seizure Detection Algorithm}
\label{sec:Method} 
\begin{figure*}

\begin{minipage}[b]{\linewidth}
  \centering
  \centerline{\includegraphics[width=\linewidth]{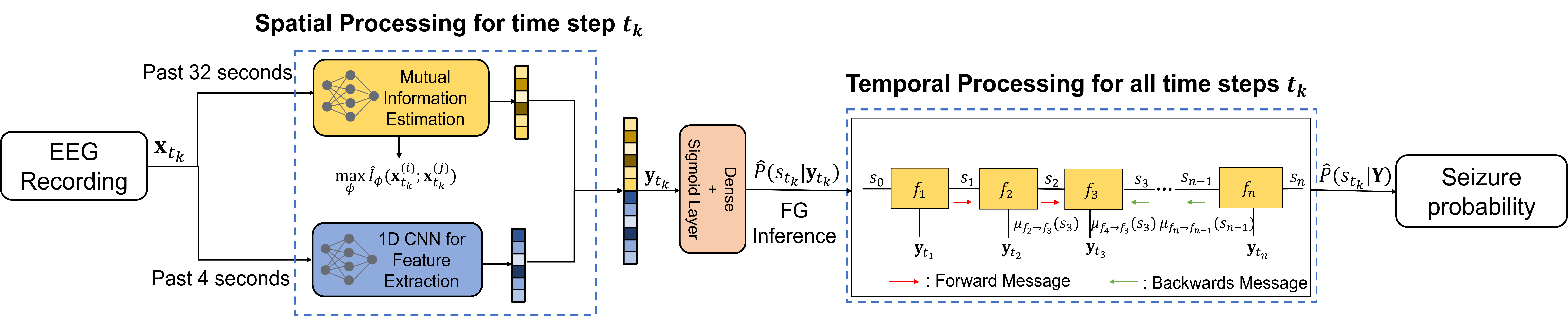}}
\end{minipage}
\caption{\ac{mical} illustration. 
}
\label{fig:method}
\end{figure*}

In this section, we present the proposed \ac{mical} algorithm. \ac{mical} is comprised of three main components:  neural \ac{mi} estimator quantifying the instantaneous dependence between different channels at each \ac{eeg} block to capture the inter-channel correlation (see Subsection~\ref{subsec:MIest}); a 1D \ac{cnn} which generates a latent representation of the raw \ac{eeg} block plus a soft estimate of the seizure state using the joint features from the 1D-CNN and the neural \ac{mi} estimator (see Subsection~\ref{subsec:1dCNN}); and factor graph inference utilizing the soft estimates as learned function nodes to incorporate temporal correlation (see Subsection~\ref{subsec:FGinference}). A high-level illustration of the flow of \ac{mical} is depicted in Fig.~\ref{fig:method}.

\subsection{Neural Mutual Information Estimation}
\label{subsec:MIest}
\ac{mi} is a measure of the statistical dependence between two random variables. While cross-correlation measures linear dependence, \ac{mi} can capture higher-order statistical dependence~\cite{kinney2014equitability}, and is thus able to capture nonlinear relationship between signals, which is likely to exist between \ac{eeg} signals during seizure~\cite{quintero2016new, jemal2021study}. The \ac{mi} between the random variables $x_1, x_2$ taking values in $\mathcal{X}\times \mathcal{X}$ with a joint distribution $P_{{X_1}X_2}$ and marginals $P_{X_1}$ and $P_{X_2}$  is defined as   
\begin{equation}
  \label{eqn:kl diverge}
I(X_1; X_2) = D_{\rm KL}(P_{{X_1}X_2}\Vert P_{X_1} P_{X_2}),
\end{equation} 
where $D_{\rm KL}$ is the \ac{kl} divergence.

Using \eqref{eqn:kl diverge} to compute \ac{mi} as a measure of statistical dependence for \ac{eeg} samples, with unknwon probability distributions is a challenging task~\cite{paninski2003estimation}. 
To address this issue, it was recently shown that neural networks can be trained to estimate \ac{mi}, based on the Donsker-Varadhan representation
\begin{align}
  D_{KL}(P_{{X_1}X_2}\Vert P_{X_1} P_{X_2}) = & \sup_{T:\mathcal{X}\times \mathcal{X} \mapsto \mathbb{R}} 
  \mathbb{E}_{P_{{X_1}X_2}}\!\left[T(x_1, x_2)\right]\! \notag \\ & \qquad \quad -\!\log\left(\mathbb{E}_{P_{X_1} P_{X_2}}\!\left[e^{T(x_1, x_2)}\right]\right),
  \label{eqn:dual kl diverge}
\end{align}
where a neural model with parameters $\boldsymbol{\phi}$ denoted by $T_{\boldsymbol{\phi}}$ is used to represent the function $T$ in \eqref{eqn:dual kl diverge}. To maximize the right hand side of~\eqref{eqn:dual kl diverge} gradient descent can be used to find the maximizing set of parameters $\boldsymbol{\phi}$ \cite{belghazi2018mutual}.
To overcome the limitations imposed by the estimation variance, \cite{song2019understanding} proposed the Smoothed \ac{mi} Lower-bound Estimator (SMILE), which learns to estimate \ac{mi} by training a neural network to maximize the objective function
\begin{align}
\hat{I}_{\boldsymbol{\phi}}(X_1; X_2) = &  ~\mathbb{E}_{P_{{X_1}X_2}}\left[T_{\boldsymbol{\phi}}(x_1, x_2)\right]\notag \\ 
& \qquad -\log\mathbb{E}_{P_{X_1} P_{X_2}}\left[{\rm clip}(e^{T_{\boldsymbol{\phi}}(x_1, x_2)},e^{-\tau},e^{\tau})\right],
  \label{eqn:smile}
\end{align}
where ${\rm clip}(v,l,u):=\max(\min(v,u),l)$ and $\tau$ is a hyperparameter. The resulting neural estimator was shown to learn to reliably predict \ac{mi} under various distributions.

\begin{figure} 
  \centering
  \centerline{\includegraphics[width=\columnwidth]{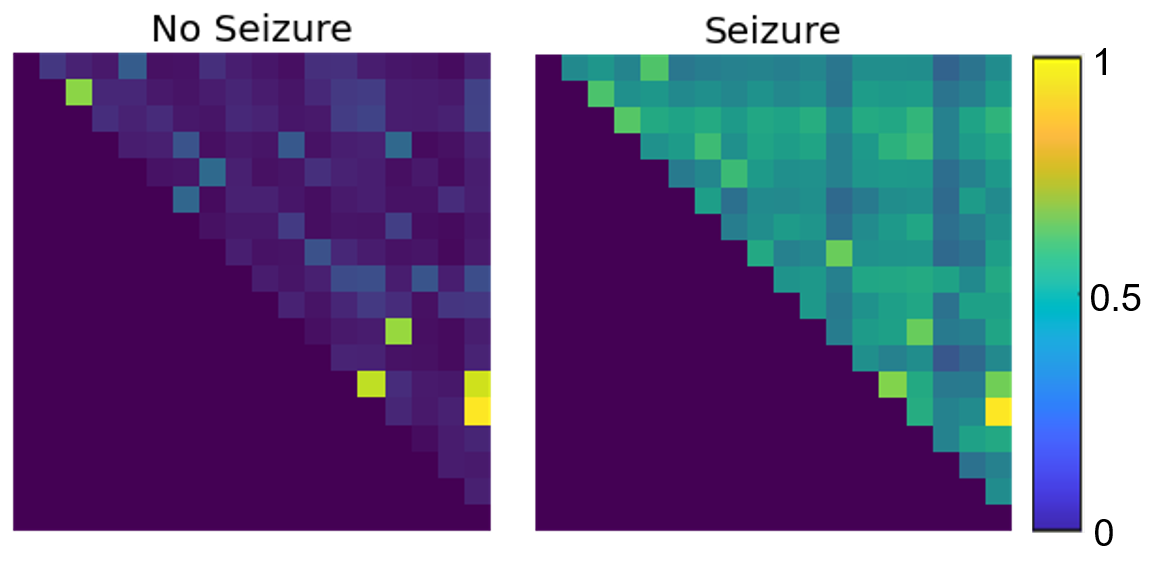}} 
\caption{Neural \ac{mi} estimation for seizure and non-seizure. 
}
\label{fig:mi result}
\end{figure}

In \ac{mical}, we apply SMILE to estimate $I(\boldsymbol{x}_{t}^{(i)};\boldsymbol{x}_{t}^{(j)})$ at each block $t$ for each channel pair $i,j$. Since \ac{mi} is symmetric, i.e., $I(\boldsymbol{x}_{t}^{(i)};\boldsymbol{x}_{t}^{(j)}) = I(\boldsymbol{x}_{t}^{(j)};\boldsymbol{x}_{t}^{(i)})$, we only estimate the \ac{mi} for $j > i$.  We set $T_{\boldsymbol{\phi}}$ to be a fully-connected network with two hidden layers and ReLU activations, and train it with $\tau = 0.9$ in the objective \eqref{eqn:smile}. The numerical results from neural estimator satisfy the underlying hypothesis of high correlation among recordings during seizure state. This is illustrated in Fig.~\ref{fig:mi result}, where  it is observed that the trained estimator outputs higher \ac{mi} values during seizure compared to non-seizure blocks.



\subsection{1D CNN}
\label{subsec:1dCNN}
In parallel to the neural \ac{mi} estimator, each raw \ac{eeg} signals block is also processed using a dedicated 1D \ac{cnn} to extract relevant features from the block. The resulting vector $\boldsymbol{y}_t$, representing the stacking of these extracted features and the estimated \ac{mi} at \ac{eeg} block $t$, is used to produce a probabilistic estimate of the presence of a seizure. We develop a 1D \ac{cnn} architecture to extract meaningful features from raw \ac{eeg} signals and combine these features with the \ac{mi} estimation results. Compared to 2D \ac{cnn}s; specifically, the baseline model proposed by Boonyakitanont et al.~\cite{boonyakitanont_comparison_2019}, 1D \ac{cnn} can evaluate all EEG channels at a given time instance, but in 2D \ac{cnn}s only the channel indexes that are close together are processed together. 

To have a comparable configuration with the baseline model, we use the same number of filters. Unlike previous studies, we design the kernel size such that our 1D \ac{cnn} will have a high receptive field of 1 second of the recording, compared to approximately 33 ms in prior works. This feature of the architecture leads to capturing low-frequency components of the signals and long-term temporal correlation within the 4-second blocks in \ac{eeg} signals. 
The details of the proposed \ac{cnn} model is shown in Fig~\ref{fig:method}. 

\subsection{Factor Graph Inference}
\label{subsec:FGinference}
The resulting seizure state probability using only the \ac{mi} estimates and the features from 1D CNN does not exploit the presence of temporal correlation. Therefore, as proposed in \cite{shlezinger2020inference} for sleep state tracking, we exploit the presence of temporal correlation by utilizing the block-wise soft decisions not for prediction, but as learned function nodes in a factor graph.
We incorporate temporal correlation by assuming that the relationship between the extracted features $\boldsymbol{y}_1,\ldots, \boldsymbol{y}_n$ and the underlying seizure state $s_1,\ldots, s_n$ can be represented as an \ac{hmm}. Similar modelling was shown to faithfully capture the temporal statistics in \ac{eeg}-based seizure detection \cite{lee_classification_2018}. For such models, one can compute the \ac{map} rule with linear complexity using sum-product inference over the resulting factor graph, as described in Subsection~\ref{subsec:Factor}. However, to evaluate the messages \eqref{eqn:ForwardPass}-\eqref{eqn:BackwardPass}, one must be able to compute the function nodes $\{f_k\}_{k=1}^n$, given by
\begin{equation}
\label{eqn:functionnode}
    f_k(s_{t_k},s_{t_{k-1}}) = P(s_{t_k}|s_{t_{k-1}}) P(\boldsymbol{y}_{t_k} | s_{t_k}).
\end{equation}

In \ac{mical}, we utilize the block-wise soft decisions as estimates of the conditional distribution $P(\boldsymbol{y}_{t_k} | s_{t_k})$. The transition probability  $P(s_{t_k}|s_{t_{k-1}})$, which is essentially comprised of two values, can be obtained from histogram, or manually tuned as we do in our numerical study in Section~\ref{sec:Sims}. 
The obtained marginal distributions \eqref{eqn:msgs} are compared to a pre-defined threshold $T$ for detection. The resulting seizure detection algorithm is summarized as Algorithm~\ref{alg:Algo1}.

 \begin{algorithm}[h]
	\caption{\ac{mical} seizure detection}
	\label{alg:Algo1}
	\textbf{Inputs:}{ SMILE and 1D CNN networks, estimated  $P(s_{t_k}|s_{t_{k-1}})$,  \ac{eeg} signals  $\boldsymbol{X}$,  threshold $T$} \\ 
	\nonl\textbf{Feature extraction:}\\
	\For{$k = 1,\ldots n$}{
	  Apply SMILE to estimate $\hat{I}_{\boldsymbol{\phi}}(\boldsymbol{x}_{t_k}^{(i)};\boldsymbol{x}_{t_k}^{(j)})$, $j > i$\;
	  Apply 1D CNN to obtain combined features $\boldsymbol{y}_{t_k}$ and obtain soft decision\;
	 } 
	\nonl\textbf{Factor graph inference:}\\
	  Compute $\{f_k\}$ from soft decisions via \eqref{eqn:functionnode}\;
	  \For{$k = 1,\ldots n$}{
	  Compute $\mu_{f_{t_k}\rightarrow s_k}(\{0,1\})$ via \eqref{eqn:ForwardPass}\;
	  Compute $\mu_{f_{t_{n-k+2}}\rightarrow s_{n-k+1}}(\{0,1\})$ via \eqref{eqn:BackwardPass}\;
	 }
	Detect seizure at $t_k$ if $\mu_{f_{t_k}\rightarrow s_k}(1)\mu_{f_{t_{k+1}}\rightarrow s_k}(1) > T$.
\end{algorithm}

\section{Numerical Results}
\label{sec:Sims}
We evaluate \ac{mical}\footnote{\href{https://github.com/bsalafia/CNN-Aided-Factor-Graphs-with-Estimated-Mutual-Information-Features-for-Seizure-Detection-MICAL.git}{The source code and hyper-parameters can be found on GitHub}.} using the CHB-MIT dataset~\cite{goldberger_physiobank_2000}. The data is comprised of scalp \ac{eeg} recordings from 24 pediatric subjects with intractable seizures, 
sampled at a frequency of 256 Hz where seizure start and end times are labeled. In order to balance and denoise the dataset, few simple pre-processing steps are included. Sample recordings with at least one seizure are selected and a notch filter is applied to remove the noise from power line. Due to the short seizure duration, for each recording file, we reduce non-seizure samples to 10 times before and 10 times after seizure time. Therefore, for every second of seizure data, there are 20 seconds of non-seizure data. The seizures are estimated for every second. To estimate the probability of seizure over the $t$-th second, the past 32 seconds of recording is used to solve the optimization that estimates \ac{mi}. This window size has demonstrated the best results over the dataset. The past 4 seconds of recording is used as input to the 1D \ac{cnn} for estimating the features. The value of 4 seconds is selected to satisfy a good trade-off between the number of samples in a  block and the stationarity of the observed signals over a block. 
In our experiments, seven models are used for comparison. The 2D~\ac{cnn} used in~\cite{boonyakitanont_comparison_2019} and spectrogram detector of \cite{jana_1d-cnn-spectrogram_2020} as two baseline models since they reported the best results compared to prior works. For \ac{mical}, we tune the transition probability to $P(s_{t_k}=1|s_{t_{k-1}}=1) = 89.54\%$ and $P(s_{t_k}=1|s_{t_{k-1}}=0) = 17.90\%$.
To evaluate the contribution of each individual component of \ac{mical}, we conduct a complete ablation study. We predict seizure probability based solely on input block through 1D~\ac{cnn} features as well as combined features from \ac{mi} estimator and \ac{cnn}. We also add two different structures, including GRU cells and factor graph to the 1D \ac{cnn} features to exploit temporal correlation without incorporating the inter-channel correlation. All detectors use decision threshold of $T = 0.5$. 

For considering variability among patients, a 6-fold leave-4-patients-out evaluation is conducted. To examine the performance of the proposed hybrid algorithm, three metrics are measured: area under ROC curve (AUC-ROC) which shows the capability to distinguish between seizure and non-seizure samples, area under precision recall curve (AUC-PR) that is the indicative of success and failure rates, and F1 score representing the harmonic mean between precision and recall.

The results for three performance measures are summarized in Table~\ref{tab:summary}. The represented values for all metrics show the average across 6 folds. As presented in Table~\ref{tab:summary}, the 1D~\ac{cnn} used by \ac{mical} achieves almost 5\% improvement compared with the baseline models, specifically for AUC-ROC and AUC-PR. As indicated, considering only temporal or inter-channel correlation has no significant effect on the model performance. Therefore, the incorporation of \ac{mi} estimation and factor graph inference by \ac{mical}  yields  the highest performance measures, $83.8\%$ and $50.38\%$ for AUC-ROC and AUC-PR, respectively and $93.42\%$ for F1 score. The results indicate that our algorithm admits the hypothesis of existing high correlation among signals during seizure states. Furthermore, exploiting temporal correlations in a principled manner through factor graphs is shown to facilitate learning an accurate detector, compared to using a black-box \acp{rnn}, at a much reduced computational complexity.

\begin{table}
\begin{center}
{\footnotesize
\begin{tabular}{|c|c c c|}\hline &  AUC-ROC &  AUC-PR &  F1 score \\
\hline
2D~\ac{cnn}~\cite{boonyakitanont_comparison_2019}& $77.81\pm0.08$ & $37\pm0.17$ & $88.3\pm0.03$ \\
\hline
Spectrogram~\cite{jana_1d-cnn-spectrogram_2020}&$75.4 \pm 0.11$ &$37.65 \pm 0.10$ & $92.77 \pm 0.03$ \\
\hline
1D CNN & $82.12\pm0.04$ & $42.23\pm0.12$ & $91.47\pm0.02$ \\
\hline
1D~\ac{cnn}-GRU &  $82.28\pm0.03$ &  $44.43\pm0.10$&  $90.42\pm0.06$ \\
\hline
1D CNN-FG & $83.15\pm0.05$ & $44.50\pm0.12$ & $92.35\pm0.02$ \\
\hline
1D CNN-SMILE & $83.10\pm0.04$ & $48.50\pm0.11$ & $92.47\pm0.01$ \\
\hline
\bf{\ac{mical}} & $\boldsymbol{83.8\pm0.04}$ & $\boldsymbol{50.38\pm0.13}$ & $\boldsymbol{93.42\pm0.01}$ \\
\hline
\end{tabular}
}
\end{center}
\vspace{-0.6cm}
\caption{Summary of results}
\label{tab:summary}
\end{table}

\section{Conclusion}
\label{sec:Conclusions}
We proposed \ac{mical}, which is a data-driven \ac{eeg}-based seizure detector designed to exploit both inter-channel and temporal correlations. For this, \ac{mical} estimates the \ac{mi} between each pair of \ac{eeg} channels to capture the non-linear correlation among recordings observed during seizure times. The estimated \ac{mi} is combined with a carefully designed 1D \ac{cnn} to provide a soft estimate for each signal block. Instead of using these features for prediction, they are utilized to evaluate the function nodes of an underlying factor graph, allowing it to infer at linear complexity while  exploiting temporal features between \ac{eeg} blocks. 
We demonstrate that \ac{mical} achieves notable improved performance compared to previously proposed methods.

\section*{Acknowledgements}
This work is supported by Discovery Grant from the
Natural Sciences and Engineering Research Council of Canada
(NSERC), grant number RGPIN-2020-04926, and Canada Foundation for Innovation (CFI), John R. Evans Leader Fund, grant number 39767.

\bibliographystyle{ieeetr}

\bibliography{IEEEabrv, Ref}

\end{document}